\documentclass[12pt]{article}
\usepackage{amsmath,amssymb,yfonts,esint}
\usepackage{wrapfig}
\usepackage{floatflt}
\usepackage{mathrsfs}

\usepackage[pdftex]{color,graphicx}

\usepackage{dsfont}
\newcommand{\be}{\begin{equation}}
\newcommand{\ba}{\begin{eqnarray}}
\newcommand{\ea}{\end{eqnarray}}
\newcommand{\nn}{\nonumber}

\def\d{\delta}
\def\e{\epsilon}

\def\k{\kappa}

\def\m{\mu}

\def\t{\tau}

\def\G{\Gamma}

\def\ca{{\cal A}}
\def\cb{{\cal B}}

\def\ch{{\cal H}}

\def\co{{\cal O}}

\def\cs{{\cal S}}

\newcommand{\pa}{\partial}

\newcommand{\bbC}{{\Bbb C}}

\fontfamily{yfrak}

\begin{document}

\vskip 15mm

\begin{center}

{\Large\bfseries Quantum Gravity coupled to Matter via\\[1ex] Noncommutative Geometry}

\vskip 4ex

Johannes \textsc{Aastrup}$\,^{a}$\footnote{email: \texttt{aastrup@uni-math.gwdg.de}},
Jesper M\o ller \textsc{Grimstrup}\,$^{b}$\footnote{email: \texttt{grimstrup@nbi.dk}}\\ \& Mario \textsc{Paschke}$^{c}$\footnote{email: \texttt{mpaschke@spassundwissenschaft.de}} 

\vskip 3ex  

$^{a}\,$\textit{Mathematisches Institut, Georg-August-Universit\"at G\"ottingen,\\ Bunsenstrasse 3, 
D-37073 G\"ottingen, Germany}
\\[3ex]
$^{b}\,$\textit{The Niels Bohr Institute, University of Copenhagen, \\Blegdamsvej 17, DK-2100 Copenhagen, Denmark}\\[3ex]

$^{c}\,$\textit{Spa{\ss} \& Wissenschaft, K\"otterstrasse 14b, D-48157 M\"unster, Germany}
%\\[3ex]
%$^{c}$ \textit{Mathematical Institute, University of Copenhagen,\\ Universitetsparken 5, DK-2100 Copenhagen, Denmark}
\end{center}

\vskip 3ex

\begin{abstract}

We show that the principal part of the Dirac Hamiltonian in 3+1 dimensions emerges in a semi-classical approximation from a construction which encodes the kinematics of quantum gravity.
The construction is a spectral triple over a configuration space of connections. It involves an algebra of holonomy loops represented as bounded operators on a separable Hilbert space and a Dirac type operator. Semi-classical states, which involve an averaging over points at which the product between loops is defined, are constructed and it is shown that the Dirac Hamiltonian emerges as the expectation value of the Dirac type operator on these states in a semi-classical approximation. %This note concludes the analysis presented in arXiv:1003.3802.
\end{abstract}

\newpage

\newpage

\section{Introduction}

In a recent series of papers \cite{AGNP1,Aastrup:2009dy,Aastrup:2010kb} it was shown that the principal part of the Dirac Hamiltonian in 3+1 dimensions emerges naturally from a construction of quantum gravity in a semi-classical limit. In this paper we resolve some important details which were left open in \cite{Aastrup:2010kb}. 

The construction from which the Dirac Hamiltonian emerges is a spectral triple over a configuration space of connections \cite{AGN3,AGN1,AGN2,Aastrup:2005yk}. Via Ashtekar variables \cite{Ashtekar:1986yd,Ashtekar:1987gu} this configuration space is related to gravity. Since a spectral triple is the basic ingredient in a noncommutative geometry this construction represents a metric construction over a space of connections. More specifically, the triple involves an algebra of holonomy loops represented as bounded operators on a separable Hilbert space on which a Dirac type operator acts. The latter resembles a functional derivation operator. The spectral triple is constructed on a projective system of ordered cubic lattices.

In \cite{AGNP1,AGN2} it was shown that the interaction between the algebra of holonomy loops and the Dirac type operator reproduces the structure of the Poisson bracket of General Relativity when formulated in terms of Ashtekar variables. This means that the spectral triple construction encodes the kinematics of quantum gravity. 

Then, in  \cite{AGNP1}, certain semi-classical states were constructed on which the principal part of the Dirac Hamiltonian in 3+1 dimensions emerges from the expectation value of the abstract Dirac type operator. However, in  \cite{AGNP1} no explanation as to why the semi-classical states had their particular form was found.  This question was addressed in \cite{Aastrup:2010kb} where a somewhat modified set of semi-classical states were found. These states have the particular property that they evade a certain dependency on a basepoint which turns up in the algebra of holonomy loops. These new states lead, however, to a classical expression which only formally equals the Dirac Hamiltonian.

In this short paper we show that a minor modification of the argumentation in \cite{Aastrup:2010kb} does lead to the emergence of the principal part of the Dirac Hamiltonian in a semiclassical limit.  \\

The paper is organized as follows: sections 2 - 5 give a review of known results. In section 2 we introduce a noncommutative $\star$-algebra of holonomy loops and a spectral triple construction over this algebra. The underlying space is an infinite dimensional space of generalized connections and the construction is based on an infinite system of nested lattices. In section 3 we link this construction to canonical quantum gravity in the sense that the spectral triple construction captures the kinematics of quantum gravity. Section 4 is concerned with the construction of states which in a certain sense smear the basepoint, at which the product between loops in the algebra is defined, over the infinite lattice. These states are then combined with coherent states, which we introduce in section 5, to form semiclassical states which, in section 6, are shown to give a Dirac operator in 3 dimensions in a semiclassical limit. In section 7 we show that a minor generalization of these states entail the principal part of the Dirac Hamiltonian in 3+1 dimensions in the same semi-classical limit.

\section{The spectral triple construction}

\begin{figure}[t]
\begin{center}
\resizebox{!}{2cm}{
 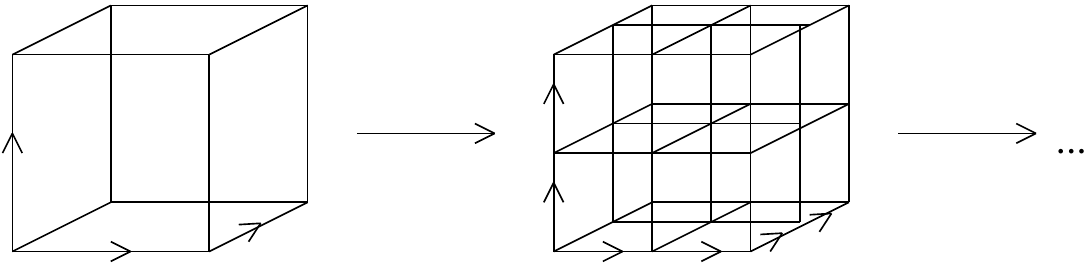}
\end{center}
\caption{\it an infinite system of nested, oriented, cubic lattices. Each lattice $\G_n$ is a symmetric subdivision of the lattice $\G_{n-1}$.}
\end{figure}

Let $\cs=\{\G_n\}_{n\in\mathbb{N}_+}$ be an ordered system of 3-dimensional, finite, cubic, oriented lattices where $\G_n$ is the symmetric subdivision of $\G_{n-1}$ as depicted in figure 1. 
These lattices can be viewed either as abstract combinatorial graphs or, as we will do in the subsequent sections, in terms of a cubulation of a compact 3-manifold $\Sigma$.

Consider first a single lattice $\G$ in $\cs$. Denote by $\{x_i\}$ and $\{l_j\}$ the vertices and edges in $\G$ where $l_j$ should be understood as a map
$$
l_j:\{0,1\}\rightarrow \{x_i\}
$$
which connects adjacent vertices $l_j(0)$ and $l_j(1)$ in $\G$. Denote by 
$$l_i^*(t)=l_i(1-t)
$$
the reversed edge.
 Let $G$ be a compact Lie-group and let $\nabla$ be a map which assigns an element in $G$ to each edge $l_j$
$$
\nabla: l_j\rightarrow   g_j \in G \;,
$$
satisfying
$$
\nabla(l_i^*)=\nabla(l_i)^{-1}\;.
$$
Denote by $\ca_{\G}$ the space of such maps. Note that 
$$
\ca_{\G}\simeq G^{n(\G)}\;,
$$
where $n(\G)$ is the number of edges in $\G$. Choose a vertex $x_0$ in $\G$ which is called the basepoint. 
A sequence of edges $\{l_{i_1},\ldots,l_{i_n}\}$, where 
$$
l_{i_k}(1)= l_{i_{k+1}}(0)\;,\quad   l_{i_1}(0)=l_{i_n}(1)=x_0\;,
$$ 
is called a loop (based in $x_0$) and a product between two loops $L_1=\{l_{i_1},\ldots,l_{i_n}\}$ and $L_2=\{l_{j_1},\ldots,l_{j_m}\}$ is given by gluing them at the basepoint $x_0$
$$
L_1 L_2 =\{l_{i_1},\ldots,l_{i_n},l_{j_1}\ldots,l_{j_m}\}\;.
$$
Two loops are considered equal if they differ by a trivial backtracking
$$
\{\ldots,l_{i_{k-1}}, l_{i_k},l_{i_k}^*,l_{i_{k+1}}\ldots\} = \{\ldots,l_{i_{k-1}}, l_{i_{k+1}}\ldots \}\;,
$$
and an involution of a loop $L$ is given by its reversal 
$$
L^* = \{l^*_{i_n},\ldots,l_{i_1}^*\}\;,
$$
which satisfies $(L^*)^*=L$ and $(L_1 L_2)^* = L_2^*  L_1^*$.
This set of loops in $\G$ form a group. Consider now formal, finite series of loops in $\G$
$$
a = \sum_k a_k L_k\;,\quad a_k\in\mathbb{C}\;,
$$
with the obvious product and involution. These formal series form a $\star$-algebra, which we denote by $\cb_{x_0,\G}$, and which has a natural norm given by
\[
\| a \|= \sup_{\nabla\in\ca_{\G}}\|\sum a_k \nabla(L_k)  \|_G\;.
\]
Here $\|\cdot \|_G$ is the matrix norm in $G$ and 
\[
\nabla(L)= \nabla({l_{i_1}})\cdot\nabla(l_{i_2})\cdot\ldots\cdot\nabla(l_{i_n})\;
\]
with $ L=\{  l_{i_1},l_{i_2}\ldots l_{l_n}  \}$.
The algebra $\cb_{x_0,\G}$ is naturally represented as bounded operators in the Hilbert space
$$
\ch'_\G =L^2(\ca_{\G},\mathbb{C}^m)\;,
$$
where $L^2$ is with respect to the Haar measure on $G$ and where $m$ is the size of a matrix representation of $G$, by
\begin{equation}
h_L\cdot \xi(\nabla) := \nabla(L) \xi(\nabla)\;,\quad \xi \in  \ch'_{\G}\;.
\label{representation}
\end{equation}
Here $\nabla(L)$ acts by matrix multiplication on the factor $\mathbb{C}^m$. 
Notice that this representation of $\cb_{x_0,\G}$ need not be faithful.

The spectral triple construction presented here works for a general compact Lie group \cite{Lai:2010ig}. However, we shall in the following assume that $G=SU(2)$ and choose its fundamental representation, i.e. m=2.

Let us now return to the full system of graphs $\cs$. With the above construction we have to each graph $\G_n$ in $\cs$ an associated space $\ca_{\G_n}$ and an algebra $\cb_{x_0,\G_n}$ represented in a Hilbert space $\ch'_{\G_n}$ of states over $\ca_{\G_n}$. We take the appropriate projective and inductive limits
$$
\overline{\ca}:=\lim_{\stackrel{\G_n\in\cs}{\leftarrow}}\ca_{\G_n}\;,\quad   \ch' := \lim_{\stackrel{\G_n\in\cs}{\rightarrow}}\ch'_{\G_n}\;,\quad
\cb_{x_0}=  \lim_{\stackrel{\G_n\in\cs}{\rightarrow}}\ \cb_{x_0,\G_n}\;.
$$
Thus, the algebra $\cb_{x_0}$ is generated by loops running in the infinite lattice $\cs$ represented as bounded operators on the separable Hilbert space $\ch'$. %The entire construction lives on the infinite dimensional space $\overline{\ca}$.

To obtain a spectral triple construction over $\overline{\ca}$ we need to construct a Dirac type operator $D$ which interacts with the algebra $\cb_{x_0}$. It turns out that the construction of such an operator requires some work. The key issue is to ensure that the Dirac type operator, which is a composition of Dirac type operators $D_\G$ acting in the intermediate Hilbert spaces $\ch'_\G$, is compatible with projections between spaces $\ca_\G$ and $\ca_{\G'}$ and corresponding embeddings between Hilbert spaces $\ch'_\G$ and $\ch'_{\G'}$. This requirement is non-trivial and restricts the choice of $D$. Here we shall simply state the result and refer to \cite{AGN2} and \cite{AGN3} for details\footnote{The construction of a Dirac type operator presented in \cite{AGN2} is different to and somewhat more involved than the construction presented in \cite{AGN3}. In the following we use the construction of  \cite{AGN3}.}.

Consider again the level of the single graph $\G_n$ which is obtained by $n$ subdivisions of the first lattice. Let $\ch_{\G_n}$ be the Hilbert space 
\begin{equation}
\ch_{\G_n}=L^2(\ca_{\G_n},Cl(T^*\ca_{\G_n})\otimes\mathbb{C}^2)\;,
\label{HI}
\end{equation}
which is obtained from $\ch_{\G_n}'$ by adding the Clifford bundle $Cl(T^*\ca_{\G_n})$.
This step requires a choice of a left and right invariant metric on $G$. Again, take the inductive limit to obtain
$$
\ch := \lim_{\stackrel{\G_n\in\cs}{\rightarrow}}\ch_{\G_n}\;.
$$
 Next, choose an orthonormal basis $\{{\bf e}_j^a\}$ of the Lie algebra of the $j$'th copy of $G$ in $\ca_{\G_n}$ and denote by
$ d_{{\bf e}_j^a}$ the corresponding left-invariant vector fields. Denote by $\bar{\bf e}_j^a$ the corresponding elements in the Clifford algebra. Consider the operator
\begin{equation}
D_{\G_n} = \sum a_j \bar{\bf e}_j^a\cdot d_{{\bf e}_j^a}\;,
\label{D}
\end{equation}
where $\cdot$ means Clifford multiplication and where $a_j$ are real parameters. In (\ref{D}) the sum runs over all the copies of $G$ assigned to edges in $\G$. However, this assignment of copies of $G$ now differs from what is described above and corresponds to a particular change of variables in $\ca_{\G_n}$.  For details we refer to  \cite{AGN3}. 

The operators $D_{\G_n}$ are compatible with embeddings of Hilbert spaces and thus entails a consistent operator $D$ acting on the limiting Hilbert space $\ch$. 

It turns out that the triple 
\begin{equation}
(\cb_{x_0},\ch,D)
\label{spec}
\end{equation} 
satisfies the requirements of a {\it semi-finite spectral triple } whenever the infinite sequence $\{a_j\}_{j\in\mathbb{N}_+}$ approaches infinity \cite{AGN3}. To have a spectral triple means that $1)$ the resolvent $(D^2+1)^{-1}$ is compact and $2)$ that the commutator of $D$ with an element in the algebra is bounded. A {\it semi-finite} spectral triple means that the first requirement is satisfied up to a trace, which means that a certain infinite degeneracy in the spectrum of $D$ is integrated out. Here, this degeneracy comes from the action of the infinite dimensional Clifford algebra acting in $\ch$. For details see \cite{AGN2,AGN3} .

\section{The kinematics of quantum gravity}

We now turn to Ashtekars formulation of general relativity in terms of connections \cite{Ashtekar:1986yd,Ashtekar:1987gu}. This formulation takes its outset in the ADM formulation and is thus based on a foliation of space-time in terms of a family of space-like 3-manifolds $\Sigma_t$. The Ashtekar variables consist\footnote{In fact, the original Ashtekar connection is a complex $SU(2)$ connection. The real case corresponds either to the Euclidean setup or to a formulation where the constraints have a more complicated form. Here we shall ignore this important issue and stick with the real connection.} of a $SU(2)$ connection $A_m^a$ and a densitized frame field $E^n_b$. 
%Here, $\{m,n,...\}$ are spatial indices and $\{a,b,...\}$ label the generators of $\mathfrak{su}(2)$. 
These variables come with a Poisson bracket
$$
\{A_n^a(x),E_b^m\} = \k\delta^a_b\d_n^m \d^3(x-y)\;,
$$
where $\k$ is the gravitational constant. In addition to this there is a set of three constraints, the Hamilton,  Diffeomorphism and Gauss constraints.

Consider now an alternative set of variables which are dual to the Ashtekar variables. We trade the connection $A_n^a$ for holonomy transforms $h_l(A)$ of the connection along curves $l$ in $\Sigma_t$
 and we trade the frame field $E^{m}_a$ for fluxes $F^S_a=\int_S  \e_{mnp} E^m_a dx^n\wedge dx^p$ of the frame field through surfaces $S$ in $\Sigma_t$. The Poisson bracket of these alternative variables reads
 \begin{equation}
\{ h_l, F^S_a \} =  \pm \kappa h_{l_1}\t_a h_{l_2}\;,
\label{Poisson}
\end{equation}
where the curve $l$ is composed of $l_1$ and $l_2$ which ends and begins, respectively, at the surface $S$. The bracket vanishes if $S$ fails to intersect $l$ and the sign of the bracket is determined by the orientation of $S$ relative to $l$. These loop and flux variables play a pivotal role in loop quantum gravity \cite{AL1}. 

In \cite{AGN1,AGNP1} we have shown that the interaction between the Dirac type operator $D$ and the algebra $\cb_{x_0}$ quantizes the bracket (\ref{Poisson}). This means that the algebra $\cb_{x_0}$ can be interpreted as an algebra of holonomy loop operators, and that the left invariant vector fields $d_{{\bf e}_j^a}$ in $D$ can be interpreted as flux operators corresponding to infinitesimal surfaces sitting at the endpoints of the edges $l_j$. We shall return to this point in section 5.

Furthermore, if we view the lattices in $\cs$ as embedded in a 3-manifold $\Sigma$, and consider the space of smooth $SU(2)$ connections\footnote{For simplicity, assume that $\ca$ is a space of connections in a trivial bundle over $\Sigma$.} on $\Sigma$, then one can show that \cite{AGN2}
$$
\ca\hookrightarrow \overline{\ca}\;,
$$
which means that the completion $\overline{\ca}$ separates the configuration space $\ca$ of connections. This result mirrors a key result in loop quantum gravity based on a system of piece-wise analytic graphs in $\Sigma_t$ (see \cite{Fleischhack:2000ij} for a thorough discussion).

Thus, these results show that the spectral triple $(\cb_{x_0},\ch,D)$ carries information which captures the kinematics of quantum gravity.

\section{Dependency on the choice of basepoint}

The algebra $\cb_{x_0}$ comes with a dependency on the basepoint $x_{0}$. Had we instead chosen to work with a commutative algebra of traced holonomy loops this dependency would not show up. To see this consider first two different basepoints  $x_0$ and $x_0'$ and two corresponding algebras $\cb_{x_0}^n$ and $\cb_{x'_0}^n$ associated to these basepoints. The relationship between $\cb_{x_0}^n$ and $\cb_{x'_0}^n$ is given by
\begin{equation}
\cb_{x_0}^n = \nabla(p)   \cb_{x_0}^n \nabla(p^*)\;,
\label{backback}
\end{equation}
where $p=\{l_{i_1},l_{i_2},\ldots,l_{i_n}\}$ is a path in $\G_n$ which connects $x_0$ and $x_0'$. If we take the trace in (\ref{backback}) the conjugation with $\nabla(p)$ vanishes due to the cyclicity of the trace and thus the traced algebra is independent on the choice of basepoint. %Thus, it is the non-tracial part of the algebra which encodes the dependency on the basepoint.

In the following we will identify states in $\ch_{\G_n}$ which show a certain degree of independency on the choice of basepoint. 
To introduce these states consider first
the commutator between $D_{\G_n}$ and $\nabla(l_i)$, which we calculate to 
\begin{equation}
[D_{\G_n},\nabla(l_i)] =\sum_a   a_i  {\bf e}_i^a g_i \sigma^a =:-  a_i \mathrm{i} \sqrt{3} V_i\;.
\label{commu}
\end{equation}
Notice that the operator $V_i$ is not unitary. To obtain a unitary operator we add an extra term to $V_i$
$$
U_i:=  \frac{\mathrm{i} }{2}\left(   {\bf e}_i^a g_i \sigma^a + \mathrm{i} {\bf e}_i^1{\bf e}_i^2{\bf e}_i^3 g_i\right) 
$$
and check that $U_i^*U_i=U_i U_i^*=\mathds{1}_2$.
Given an element $L\in\cb_{x_0,\G_n}$ we compute
$$
\mbox{Tr}_{\tiny Cl}\left( V_i L V_i^*\right)  =  L_0  -\frac{1}{3} g_i L^a\sigma^a g_i^* \;,\quad  \mbox{Tr}_{\tiny Cl}\left(  U_i L U_i^* \right) =  L_0 \;,
$$
where we write $L= L_0 + L^a \sigma^a$ and where $\mbox{Tr}_{\tiny Cl}$ denotes the trace over the Clifford algebra. Thus, conjugating with either $V_i$ and $U_i$ singles out the trace of $L$. In the first case the non-tracial part is suppressed with a factor $\frac{1}{3}$, in the second case the non-tracial part of $L$ is absent altogether.
Next, let again $p=\{ l_{i_1}, l_{i_2},\ldots , l_{i_n}  \}$ be a path in $\G_n$ and define two associated operators by
$$
V_p := V_{i_1} V_{i_2} \ldots V_{i_n}   \;,\qquad U_p := U_{i_1}U_{i_2} \ldots U_{i_n}\;.
$$
Both $V_p$ and $U_p$ are easily seen to form two families of mutually orthogonal operators labelled by paths in $\G_n$
$$
\mbox{Tr}_{\tiny Cl}\left(  V_p^* V_{p'}\right) = \d_{p,p'}\;,\qquad \mbox{Tr}_{\tiny Cl}\left(  U_p^* U_{p'}\right) = \d_{p,p'}\;,
$$
due to the presence of the Clifford algebra elements in $V_p$ and $U_p$. Here $\d_{p,p'}$ equals one when the paths $p$ and $p'$ are identical and zero otherwise. 
%Notice, due to (\ref{commu}), that these operators resemble $n$-forms (where???).

Consider now states in $\ch_{\G_n}$ of the form
\begin{equation}
\Psi_{n}(\psi) = 2^{-3n}\sum_i U_{p_i}\psi(x_i)\;,
\label{sss}
\end{equation}
where the sum runs over vertices\footnote{excluding the basepoint.} $x_i$ in $\G_{n}$ where a path $p_i$ connects the basepoint $x_0$ with a vertex $x_i$. Also, $\psi(x_i)$ denotes an element in $\mathbb{C}^2$ associated to the vertex $x_i$. Later, upon taking a certain continuum limit, $\psi(x_i)$ will be seen to represent a (Weyl) spinor degree of freedom at the point $x_i$.

$\Psi_n$ is a state in $\ch_{\G_n}$ which does not depend on the choice of basepoint $x_0$ in the sense that the expectation value of an element $L$ in $\cb_{x_0,\G_n}$ on this state will depend only on the trace of $L$
\begin{equation}
\langle \Psi_{n} \vert L \vert \Psi_{n}\rangle =\langle \Psi_{n} \vert  \mbox{Tr}(L) \vert \Psi_{n}\rangle = \langle   \mbox{Tr}(L)\rangle\sum_i \psi^\dagger(x_i)\psi(x_i)\;.
\label{indy}
\end{equation}

The states (\ref{sss}) may alternatively be constructed using the operators $V_p$. In this case the non-tracial part of the expectation value of the algebra will only vanish upon taking the continuum limit which will be introduced in the following. However, the main results of this paper concerning the Dirac Hamiltonian holds also for such alternative states.

\section{Coherent states on $\ca_{\G_n}$}

Before we proceed we recall the results for coherent states on compact connected Lie groups that we are going to use. For simplicity we will only consider various copies of $SU(2)$. This construction uses results of Hall \cite{H1,H2}, and is inspired by the articles \cite{BT1,TW,BT2}.

First pick a point $(A_n^a,E^m_b)$ in the phase space of Ashtekar variables on a 3-manifold $\Sigma$. The states which we construct will be coherent states peaked over this point.
Consider first a single edge $l_i$ and thus one copy of $SU(2)$. Let $\{{\bf e}^a_i\}$ be a basis for $\mathfrak{su}(2)$. 
%Given $g_0$ in $SU(2)$ and given three momenta (real numbers) $E^1,E^2,E^3$
There exist families $\phi^t_{l_i}\in L^2(SU(2))$ such that
$$ \lim_{t \to 0}\langle \phi^t_{l_i}, t d_{{\bf e}^a_i}\phi_{l_i}^t \rangle=2^{-2n}\mathrm{i}E_a^m(x_{j})\;,$$
and
$$\lim_{t \to 0}\langle \phi_{l_i}^t\otimes v, \nabla(l_i)\phi_{l_i}^t\otimes v \rangle=(v,h_{l_i}(A)v)\;,$$
where $v \in \bbC^2$, and $(,)$ denotes the inner product hereon; $x_{j}$ denotes the right endpoint of $l_i$, and the index "$m$" in the $E^m_a$ refers to the direction of $l_i$. The factor $2^{-2n}$ is due to the fact that $d_{{\bf e}_j^a}$ corresponds to a flux operator with a surface determined by the lattice \cite{AGNP1}.
Corresponding statements hold for operators of the type $$f(\nabla(l_i))P(t d_{{\bf e}^1_i},t d_{{\bf e}^2_i},t d_{{\bf e}^3_i}),$$  where $P$ is a polynomial in three variables, and $f$ is a smooth function on $SU(2)$, i.e.
$$ \lim_{t \to 0}\langle \phi^t_{l_i} f(\nabla(l_i))P(t d_{{\bf e}^1_i},t d_{{\bf e}^2_i},t d_{{\bf e}^3_i}) \phi^t_{l_i} \rangle=f(h_{l_i}(A))P(\mathrm{i}E_1^m,\mathrm{i}E_2^m,\mathrm{i}E_3m)\;.$$
%This statement also carries over to symbols, i.e. functions on $T^*SU(2)$ with certain properties.

 The states $\phi^t_{l_i}$  have further important physical properties which we are however not going to use at the present stage of the analysis. Also, the precise construction of these states, in particular the choice of complexifier \cite{TW}, is irrelevant for the results presented in this paper.

Let us now consider the graph $\G_n$. We split the edges into $\{ l_i \}$, and $\{ l'_i\}$, where $\{ l_i\}$ denotes the edges appearing in the $n$'th subdivision but not in the $n-1$'th subdivision, and $\{ l'_i \}$ the rest. Let
$\phi^t_{l_i}$ be the coherent state on $SU(2)$ defined above and
%such that
%$$\lim_{t \to 0}\langle \phi^t_{l_i} \otimes v ,\nabla(l_i) \phi^t_{l_i}\otimes  v \rangle =  (v,h_{l_i}(A)v)\;,$$
%and
%$$\lim_{t \to 0}\langle \phi^t_{l_i} , td_{{\bf e}_i^a} ,\phi^t_{l_i}\rangle = 2^{-2n}\mathrm{i}E_a^m(x_{j})\;,$$
%where $v \in \bbC^2$; $x_{j}$ denotes again the right endpoint of $l_i$, and the $m$ in the $E^m_a$ refers to the direction of $l_i$. The factor $2^{-2n}$ is due to the fact that the left invariant vector field $d_{{\bf e}_j^a}$ corresponds to a flux operator with a surface determined by the lattice \cite{AGNP1}.
define the states $\phi_{l'_i}$ by
$$\lim_{t \to 0}\langle \phi^t_{l'_i} \otimes v ,\nabla(l_i) \phi^t_{l'_i}\otimes  v \rangle =  (v,h_{l'_i}(A)v)\;,$$
and
$$\lim_{t \to 0}\langle \phi^t_{l'_i} , t d_{{\bf e}_j^a} \phi^t_{l'_i}\rangle = 0\;.$$
Finally define $\phi^t_n$ to be the product of all these states as a state in $L^2(\ca_{\Gamma_n})$. These states are essentially identical to the states constructed in \cite{TW} except that they are based on cubic lattices and a particular mode of subdivision.

In the limit $n\rightarrow\infty$ these states produce the right expectation value on all loop operators in the infinite lattice.

\section{Semi-classical analysis, the Dirac operator in 3 dimensions}

The purpose of this and the subsequent section is to analyse a semi-classical approximation of the spectral triple construction (\ref{spec}).  We will show that the expectation value of $D$ on the particular states constructed in section 4 gives, in a certain limit, the Dirac operator in 3 dimensions and, with a certain modification, the Dirac Hamiltonian in $3+1$ dimensions.

 Consider first the states
\begin{equation}
\Psi^t_n(\psi) :=  \Psi_{n}(\psi)  \phi^t_n
\label{dadada}
\end{equation}
composed by the states (\ref{sss}) and the coherent states introduced in the previous section. Furthermore, we will restrict the series $\{a_j\}_{j\in\mathbb{N}_+}$ of parameters in $D$: we now require all $a_j$'s associated to edges appearing in the $k$'th subdivision but not in the $k-1$'th subdivision to be equal. With this restriction the parameters $a_j$ represent a scaling degree of freedom.

We find %(CHECK!)
\begin{eqnarray}
\langle \Psi^t_n\vert  D \vert \Psi^t_n\rangle \hspace{-2.1cm} &&
 \nn\\&=&
 \frac{a_n}{2^{5n+1}} 
 \sum_i \Big(  \psi^\dagger(x_i)  E^m_a g_{i}\sigma^a  \psi(x_{i+1}) +(g_{i} \sigma^a  {\psi}(x_{i+1}))^\dagger  E^m_a \psi(x_{i}) 
 \Big)\;,
 \label{hmmmmm}
\end{eqnarray}
where $x_i$ and $x_{i+1}$ denote start and endpoint of an edge $l_i$ and where the sum runs over edges appearing in the $n$'th but not in the $n-1$'th subdivision. Expand $g_{i}$ according to
$$
g_{i}= \mathds{1}_2 + \e A_m(x_i) +\co(\e^2)\;,
$$
where "$m$" is again the direction of the edge  $l_i$. This expansion is permitted whenever we apply the coherent states and take the continuum limit $n\rightarrow\infty$ together with the semi-classical limit $t\rightarrow 0$. Then equation (\ref{hmmmmm}) gives
\begin{equation}
\lim_{n\rightarrow\infty}\lim_{t\rightarrow 0} \langle \Psi^t_n\vert t D \vert \Psi^t_n\rangle =
%\nn\\&&\hspace{-2.5cm}
 \frac{1}{2}  \int_\Sigma d^3x  {\psi}^\dagger(x) \Big( E^m_a \nabla_m \sigma^a  + \nabla_m E^m_a  \sigma^a    \Big) \psi(x) 
 \label{hmmm}
\end{equation}
where we set $\e=2^{-n}$ and write $\nabla_m=\pa_m+ A_m$. Crucially, obtaining (\ref{hmmm}) requires that we fix the free parameters $\{a_n\}$ to $a_n= 2^{3n}$.
Thus, the expectation value of the Dirac type operator $D$ on the states (\ref{dadada}) renders, in a combination of a semi-classical limit and a continuum limit, the expectation value of an ordinary Dirac operator on a manifold $\Sigma$. 

The expression (\ref{hmmm}) is formulated with respect to a certain coordinate system which emerges from the graphs $\G_n$. Thus, the cubic lattices in $\cs$ can, in this specific limit, be interpreted as an emergent coordinate system. In particular, this coordinate system is exactly the coordinate system in which the classical Ashtekar variables are formulated\footnote{Of course, the Ashtekar variables do not depend on a particular choice of a coordinate system. However, the way they enter the analysis in this paper, they are written down with respect to a coordinate system.}. 

Notice that in order to obtain expression (\ref{hmmm}) we had to fix the parameter $a_n$. Thus, the freedom which we encountered when we constructed the Dirac type operator $D$ is eliminated in order to obtain a sensible semi-classical limit.

Notice also that $E^m_a$ is the densitized dreibein $E^m_a = \sqrt{g} e^m_a $ where $e^m_a$ is the dreibein and $g$ is the determinant of the 3-metric $g$ given by the dreibein. This means that the inverse measure $d^3x \sqrt{g}$ arises naturally in (\ref{hmmm}).

\section{The Dirac Hamiltonian}

The states (\ref{sss}) and (\ref{dadada}) are not the most general states on which the action of the algebra $\cb^n_{x_0}$ is independent of the basepoint. In this section we show that generalizations of (\ref{sss}) which still satisfy (\ref{indy}) lead naturally to the Dirac Hamiltonian in $3+1$ dimensions.

Consider the modified operators
$$
\tilde{U}_i = U_i M(i) 
$$
where $M(i)$ is an arbitrary, self-adjoint two-by-two matrix associated to the edge $l_i$. Following the recipe of section 4 we first construct the corresponding operators
\begin{equation}
\tilde{U}_p := \tilde{U}_{i_1}\tilde{U}_{i_2} \ldots \tilde{U}_{i_n}\;,
\label{Nichols}
\end{equation}
as well as the corresponding states
\begin{equation}
\tilde{\Psi}^t_n(\psi) = \left(2^{-3n}\sum_i \tilde{U}_{p_i} \psi(x_i)\right) \phi^t_n\;.
\label{dadadadi}
\end{equation}
One finds that the expectation value of $D$ on these modified states gives
\begin{eqnarray}
\lim_{n\rightarrow\infty}\lim_{t\rightarrow 0} \langle \tilde{\Psi}^t_n\vert t D \vert \tilde{\Psi}^t_n\rangle
\hspace{-3.6cm}&&\nonumber\\ &=& 
%\nn\\&&\hspace{-2.5cm}
 \frac{1}{2}  \int_\Sigma d^3x  {\psi}^\dagger(x) \Big( (N + N^b \sigma^b)(E^m_a \nabla_\m \sigma^a  + \nabla_\m E^m_a  \sigma^a )   \Big) \psi(x) \;,
 \label{hmmmdi}
\end{eqnarray}
where we wrote $M(i)$ as $N(x)+ \mathrm{i}N^a(x)\sigma^a$ with $x$ referring to the point which $l_i$ singles out in this limit. Here, $N(x)$ and $N^a(x)$ are seen to give the lapse and shift fields.  In (\ref{hmmmdi}) we have omitted certain zero-order terms. Equation (\ref{hmmmdi}) is thus seen to equal the principal part of the Dirac Hamiltonian in $3+1$ dimensions. Thus, the states (\ref{Nichols}) can be interpreted as one-particle states on which the Dirac type operator $D$ gives the Hamiltonian in the semi-classical approximation.
We refer to \cite{AGNP1} for a more detailed discussion of this interpretation.

The result (\ref{hmmmdi}) holds only if the matrices $M(i)$ are unitary, which means that the lapse and shift fields are normalized as a 4-vector.  It is possible to introduce the lapse and shift fields in a different manner. For instance, they might emerge from a polar decomposition of an element in $SL(2,\mathbb{C})$ which in turn could come from a construction which operates with two copies of $SU(2)$ assigned to each edge. This option would not put any restrictions on the lapse and shift fields. Alternatively, in \cite{Aastrup:2009dy} we suggest a different construction involving a generalized Dirac type operator. Which construction is better must be determined through a deeper understanding of the spectral triple construction.

Notice, that the coherent states $\phi^t_n$ are, in the limit where $n$ approaches infinity, no longer states in the Hilbert space $\ch$. Furthermore, in this limit combined with the semiclassical approximation, an action of the diffeomorphism group, which is clearly absent at the level of a finite graph, emerges. Thus, one may speculate what structure these coherent states generate and whether this structure carries an action of the diffeomorphism group. We suspect that this might be the true object of interest here and that the lattices simply serve as an intermediate step towards a continuum construction.

Finally, having now established the existence of one-particle states it should be determined whether the Hilbert space $\ch$ also involves many-particle states. This is the subject of present investigations.

\end{document}